\documentclass[aps,prb,floatfix,twocolumn,superscriptaddress]{revtex4}  
\usepackage{amsmath,amssymb,graphicx,bm,epsfig} 
 
\begin{document} 
 
\title{Exponential Convergence of Cellular Dynamical Mean Field
Theory: Reply to the comment by  K. Aryanpour, Th. Maier and M. 
Jarrell (cond-mat/0301460S)} 
\author{G. Biroli}
\affiliation{Service de Physique Th{\'e}orique, CEA Saclay 91191 Gif-Sur-Yvette, France}
\author{G. Kotliar} 
\affiliation{Department of Physics and Center for Material Theory, Rutgers 
University, Piscataway, NJ 08854, USA} 
\date{\today} 
\begin{abstract} 
We reply to the comment by  K. Aryanpour, Th. Maier and M. 
Jarrell  (cond-mat/0301460) on our paper (Phys. Rev. B {\bf 65} 
155112 (2002)).  We demonstrate using  general arguments and explicit 
examples that whenever the correlation length is finite, local 
observables converge exponentially fast in the cluster size, 
$L_{c}$, within Cellular Dynamical Mean Field Theory (CDMFT). 
This is a  faster rate of convergence  than  the $1/L_{c}^{2}$ 
behavior of the  Dynamical Cluster approximation (DCA) 
thus refuting the central assertion of their comment.
 
\end{abstract} 
 
\maketitle 
 
The development of  cluster extensions of dynamical mean 
field theory is an active area of research. Cluster  dynamical 
mean field theories, construct approximations to the solution 
of model Hamiltonians on the lattice in terms of the solution 
of a cluster  impurity model. Different cluster schemes should 
be viewed as different truncations  of the full quantum many body 
problem. As the  size of the cluster  tends to infinite, 
all cluster schemes approach the exact solution of the lattice model. 
A relevant question is, 
 for a given computational power, 
(which only allows the investigation of small cluster sizes) which 
truncation is closer to  the result in the  thermodynamic limit.\\ 
Our previous publication \cite{biroli} 
 investigated  and extended two cluster schemes, the cellular 
dynamical mean field theory \cite{cdmft} (CDMFT), and the 
dynamical cluster approximation \cite{dca} (DCA) by applying it to an exact solvable model  and 
concluded that CDMFT converges faster than DCA to the exact solution 
of that model. 
In their comment \cite{comment}, Aryanpour et. al introduce 
a new generalization of the original   DCA equations 
that take into account better the non-local 
interaction, and they argue that their new method 
converges faster than CDMFT to the exact solution of the model as 
the size of the cluster increases. They also comment that our  findings of 
rapid convergence of CMDFT are surprising, in the light of an earlier 
publication \cite{thomas} in which they 
concluded that CDMFT converges to the infinite cluster size limit with corrections 
of order $O (1/L_c) $ where $L_c$  is the  size of the cluster while 
DCA converges faster, with corrections of order $O (1/L_c^2) $.\\ 
In this reply to their comment  we point out that  {\it local observables }  in CDMFT 
generally 
converge exponentially  at finite temperatures, as long as the relevant  correlation 
length is finite (a situation that excludes a critical point). 
This statement persists at zero temperature in systems 
which have an energy  gap.\\ 
We  demonstrate  the exponential convergence of CDMFT in three steps. 
First we present general arguments 
in favor of  exponential  convergence of local 
observables in CDMFT  whenever the  relevant correlation length is finite. 
This is a direct consequence of the cavity construction underlying 
the method. 
In the process  we explain  why the convergence 
criterion introduced in ref \cite{thomas} and used to conclude that 
CDMFT converges as $1/L_{c}$ is  not an appropriate measure 
of convergence of local observables in CDMFT, which instead 
converge much faster than a power law in $L_{c}$. 
Then we present the numerical results for the $SU (N)$ spin chain 
studied in \cite{biroli} that agree completely with our general arguments 
and we display explicitly an example of the exponential 
convergence in this model. Finally, we discuss another simple case, the semiclassical limit 
of the Falikov-Kimball model in one dimension.  Previous work 
established \cite{olivier} that  in this limit quantum cluster 
approximations reduced to classical cluster approximations. 
This allow us to compare DCA and CDMFT in detail using simple 
analytical considerations. In particular we unveil, in an explicit 
example, that the DCA predictions for local observables converge as 
$1/L_{c}^{2}$ even when the same quantities obtained solving a finite system of 
size $L_{c}$ (with for example periodic boundary condition) 
converge exponentially fast (in $L_{c}$) to their thermodynamic limit value. 
 
Let us start with some general considerations. 
If one was able to trace out exactly all the degrees of freedom 
outside the cluster to get an exact effective action for the 
degrees of freedom inside the cluster then the translation invariance 
of the effective action 
would be broken  (degrees of freedom near the boundary of the cluster are  affected by 
the enviroment more than the bulk degrees of freedom)  
but  observables within the cluster would still be translationally invariant. 
So on very general grounds we expect that the Weiss field, (or 
hybridization function $\Delta$) 
which describes the effects of the degrees of freedom integrated out, is 
large near the boundary and small
(in fact exponentially small 
if the correlation length is finite) inside the cluster.\\ 
CDMFT is an approximate way 
to realize this cavity construction. 
It  produces a Weiss field which is large at the boundary 
and small in the bulk, but because of its approximate character it 
produces non   translation invariant observables. 
However, whenever the correlation length is 
finite:  (1) bulk quantities for a free system of 
size $L$ with, for example, free boundary conditions converge 
exponentially fast in $L$ to their thermodynamic limit (2) the CDMFT 
approximation should improve the convergence of bulk quantities, 
which as a consequence should be at least as fast as the one of the 
system with free  boundary conditions. 
Note that this is not the case of DCA that still converges as 
$1/L^{2}_{c}$ worse than the results for a  finite system with, for example, 
free boundary conditions which  would 
converge exponentially fast in $L_{c}$! We will discuss an explicit 
example of this behavior below. 
 
\begin{figure}[thb] 
\[ 
\includegraphics[width=7cm]{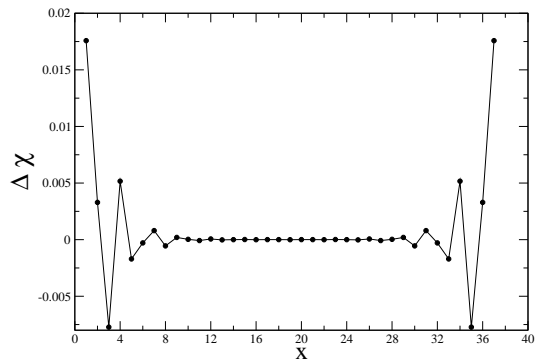} 
\] 
\caption{ 
\label{profile.eps} 
$\chi_{x}-\chi$ as a function of $x$, for $\beta $, the inverse 
temperature, equal to $4$, $L_{c}=38$ 
and $t=\mu =1$.} 
\end{figure}

Because CDMFT  breaks the translation invariance inside the cluster 
it is important to extract properly the value of local observables. 
These are well represented in the center of the cluster (see Fig. 1) , and will 
have more deviations near the boundary. The role of the cavity field 
is to try to reduce these deviations, but in approximate treatments 
there will always be errors of order one near the boundary. 
Aryanpour et al.\cite{comment} concluded the $1/L_{c}$ convergence 
by estimating the value of a local 
observable of interest  by doing a flat average (of the value of the 
local observable) over the cluster. 
Since the error is of the order one on the boundary, one 
obtains an error that dies out as the ratio surface over volume of the 
cluster, i.e. $1/L_{c}$.  On the other hand, if we extract 
this value from the center of the cluster, which is the natural 
prescription  dictated by the approximate CDMFT cavity construction, 
we obtain exponential convergence as we show below.\\ 
We now turn to the simple one dimensional $SU (N)$ chain studied 
in \cite{biroli} whose Hamiltonian is: 
\begin{eqnarray}\label{hamiltoniansimple} 
H&=&-t\sum_{i,\sigma }(f_{i,\sigma }^{\dagger}f_{i+1,\sigma } 
+f_{i+1,\sigma }^{\dagger}f_{i,\sigma })\\ 
&+&\frac{J}{2N}\sum_{i,\sigma ,\sigma '}(f_{i,\sigma }^{\dagger}f_{i,\sigma' }f_{i+1,\sigma '}^{\dagger}f_{i+1,\sigma}+f_{i+1,\sigma }^{\dagger}f_{i+1,\sigma' }f_{i,\sigma '}^{\dagger}f_{i,\sigma})\nonumber 
\end{eqnarray} 
where $f$ are fermions operators, $i=1,\dots ,L$ and $\sigma =1, \dots 
, N$ and we take the large $L$ and $N$ limits. 
This model is a generalization introduced by I. Affleck and B. Marston 
\cite{affleck} of the Hubbard-Heisenberg model where 
the SU(2) spins are replaced by SU(N) spins, the on site repulsion is 
scaled as $1/N$ and the large N limit is taken.\\ 
One can apply CDMFT and DCA to this model. Note however, that because 
the interaction is non-local, there are different possible extensions 
of usual cluster methods to this case. We extended DCA in a way based 
on the real space perspective \cite{biroli}, Aryanpour et. al introduced 
a different generalization of DCA  which takes into account better the 
non-local interactions. Our procedure is therefore not an incorrect 
application of DCA, as claimed by Aryanpour et al., 
but only a different generalization of DCA to the case of 
non-local interaction. The results of the two different generalization 
are discussed in \cite{biroli,comment} and in the following we shall 
focus on the generalization of Aryanpour et al. which has been shown to 
converge to the thermodynamic limit with an error of the order of 
$1/L_{c}^{2}$ where $L_{c}$ is the size of the cluster. 
This  rate of convergence is a general property of DCA \cite{thomas}, 
at least far from critical points.\\ 
In the following we will use $J$ as the unit of temperature 
and therefore we put $J=1$ and we rescale the hopping term $t\rightarrow t/J$. 
The thermodynamics of this model can be solved exactly 
since in the large $N$ limit the quantity $ \chi = \frac{1}{N}\sum_{\sigma }f_{i,\sigma }^{\dagger}(t)f_{i+1,\sigma}(t)$ does not fluctuate. Indeed 
(\ref{hamiltoniansimple}) reduces to a free-fermions 
Hamiltonian with a ``renormalized'' hopping term $t\rightarrow t+\chi $ 
and a self-consistent condition on $\chi $: 
\begin{equation}\label{selfsimple} 
\chi =\frac{1}{L}\sum_{k}f(\beta E_{k})\cos k, \qquad E_{k}=-2(t+\chi )\cos k +\mu 
\end{equation} 
where $\mu$ is the chemical potential, $f(\beta E_{k})$ is the Fermi function 
and $\beta $ is the inverse temperature\cite{biroli}.\\ 
DCA and CDMFT result in self-consistent eqs. for $ \chi _{x}= 
\frac{1}{N}\sum_{\sigma }<f_{x,\sigma }^{\dagger}f_{x+1,\sigma}>$ 
\cite{biroli,comment}. Because of the translation invariance of DCA the $\chi 
_{x}^{DCA}$ are independent of $x$ inside the cluster. Once self-consistency 
is achieved one can use the cluster quantity to obtain the DCA 
lattice prediction $\chi_{latt}^{DCA}$ as explained in \cite{comment}. 
This quantity converges to the exact $\chi$ with corrections of $O 
(1/L^{2}_{c})$ \cite{comment,thomas}. 
Contrary to DCA, the self-consistent CDMFT eqs. breaks the translation 
invariance inside the cluster.\\ 
In fig. \ref{profile.eps} we plot 
the error of the CDMFT prediction on $\chi$, $\chi_{x}-\chi$, as a function of $x$, where $\chi_{x}$ is the CDMFT 
solution for $\beta $, the inverse temperature, equal to $4$, $t=\mu
=1$ and $L_{c}=38$.\\

\begin{figure}[thb] 
\[ 
\includegraphics[width=7cm]{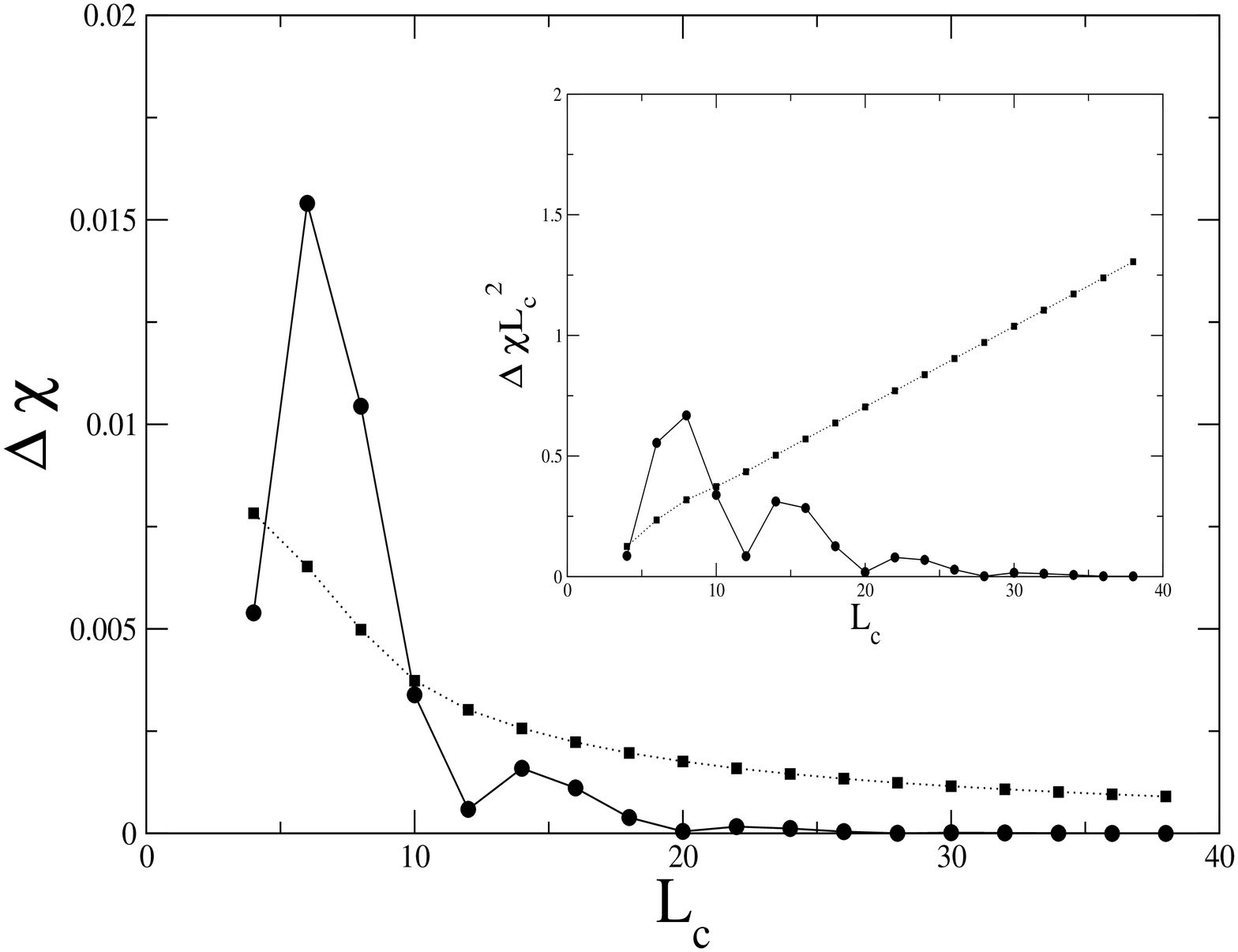} 
\] 
\caption{ 
\label{errb3.eps} 
$\Delta \chi_{bulk}$ (circles, solid line) and $\Delta \chi_{fa}$ 
(squares, dotted line) 
as a function of $L_{c}$. Inset: $\Delta \chi_{bulk}L_{c}^{2}$ 
(circles, solid line) and $\Delta \chi_{fa}L_{c}^{2}$ (squares, dotted 
line) as a function of $L_{c}$. } 
\end{figure}

\begin{figure}[thb] 
\[ 
\includegraphics[width=6cm]{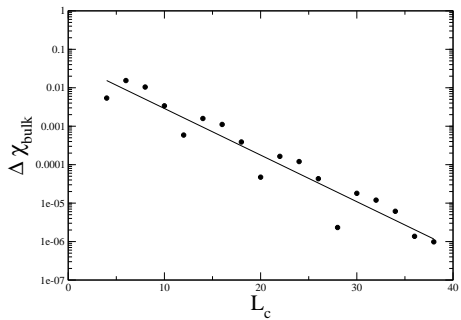} 
\] 
\caption{$\Delta \chi_{bulk}$ as a function of $L_{c}$ in a 
logarithmic scale. 
\label{errb3bulk.eps} 
} 
\end{figure}

This figure clearly shows the behavior discussed previously, 
namely   translation invariance inside the cluster  is broken: 
errors are smaller  in the bulk while they remain of the order one 
at the boundary.   Aryanpour et al. \cite{comment} concluded, by 
carrying out  a flat average $\sum_{x}\chi_{x}/ (L_{c}-1)$ over 
the cluster, that the error within CDMFT is  expected to be of the 
order $1/L_{c}$ (more generically is surface over volume, hence, 
$1/L_{c}$ also in dimension larger than one). As discussed 
above,  it is better to extract the CDMFT estimators weighting bulk 
values more than boundary values. The easiest thing to do is just 
taking the value of $\chi_{x}$ at the center of the cluster. In 
fig. \ref{errb3.eps} we compare the error obtained doing the flat 
average (square, dotted line) to the one obtained focusing on the 
bulk values (circle, solid line). This figure conveys two 
important information: first the $\Delta 
\chi_{bulk}=|\chi_{L_{c}/2}-\chi|$ is much smaller than the flat 
average one $\Delta \chi_{fa}=|\sum_{x}\chi_{x}/ (L_{c}-1)-\chi|$ 
for large values of $L_{c}$. Second, as shown in the inset, the 
error $\Delta \chi_{bulk}$ multiplied by $L_{c}^{2}$ is still 
decreasing fast as a function of $L_{c}$, i.e. $\Delta 
\chi_{bulk}$ decreases much faster than $1/L_{c}^{2}$ . Instead 
the error corresponding to the flat average leads to a straight 
line corresponding to the $1/L^{c}$ behavior discussed above and 
in \cite{comment}. Note that in this plot the DCA prediction 
would lead to a function approaching a constant when 
$L_{c}\rightarrow \infty $. 
Finally, in fig. \ref{errb3bulk.eps} we plot $\Delta \chi_{bulk}$ 
in a logarithmic scale as a function of $L_{c}$. 
The exponential convergence is manifest (the straight line is a guide 
for the eye) and in complete agreement with our general discussion
above.

Another simple example that sheds light on  
the convergence properties of CDMFT and DCA is 
the Falikov-Kimball model in the large $U$ limit. In this limit 
the system becomes a classical Ising model and the quantum cluster schemes map 
on classical cluster schemes \cite{olivier}. 
In the following we will focus on the easy one dimensional case 
in the paramagnetic phase. In this case, since the Weiss field is 
zero, doing CDMFT on a cluster of size $L_{c}$ corresponds only 
to solving a finite system of size $L_{c}$ with free boundary conditions. 
Instead DCA corresponds to solving a finite system of size $L_{c}$ 
with periodic boundary conditions and a coupling 
\[ 
J_{DCA}=J\left( \frac{\sin \frac{\pi }{2L_{c}}}{\frac{\pi}{2L_{c}}}\right)^{2} 
\] 
where $J$ is the original spin-spin coupling. So, DCA and CDMFT differ 
in the different boundary conditions but {\it also} in the fact that 
for DCA the internal spin-spin coupling is modified from its original 
value whereas for CDMFT is not. This leads to very different 
convergence properties. Let us focus for example on the prediction 
for the spin-spin correlation $C=<S_{i}S_{i+1}>$. 
The CDMFT and DCA results read: 
\[ 
C_{CDMFT}=\tanh \beta J 
\] 
\[ 
C_{DCA}=\frac{\tanh \beta J_{DCA}+ (\tanh \beta J_{DCA})^{L_{c}-1}}{1+(\tanh 
\beta J_{DCA})^{L_{c}}} 
\] 
In this case the CDMFT prediction is exact because tracing out the 
spins outside the cluster indeed leads to a zero magnetic field on the 
boundary but this is of course a peculiarity of this simple case. 
Instead, there are two types of corrections to DCA (note that 
$J_{DCA}=J+O (1/L_{c^{2}})$). There are corrections 
which die out exponentially fast as $\propto (\tanh \beta J)^{L_{c}}$. 
These are the same type of corrections that one obtains using periodic 
boundary conditions or other type of boundary conditions for a free system. 
However, there is a much larger correction coming from the first term 
in the numerator of $C_{DCA}$ which leads to 
\[ 
C_{DCA}=\tanh \beta J-\frac{\beta J\pi^{2}}{3(\cosh \beta J)^{2} (2L_{c})^{2}}+O (L_{c}^{-4}) 
\] 
Thus applying DCA to this problem one obtain a convergence as 
$1/L_{c}^{2}$ which is much worse than the exponential one 
corresponding to solving the free model with periodic boundary 
conditions.  The origin of this behavior can be 
traced to the fact that as DCA forces translation invariance 
inside the cluster the couplings are changed everywhere in the 
system by  an amount of the order $1/L_{c}^{2}$. So even if the 
correlation length is finite this error dominates 
the convergence. \\ 
While we stress the obvious advantages of CDMFT, it is also 
worthwhile to point out the aspects of the CDMFT method  (and 
cluster methods in general) that still require development. The 
lack of translation invariance of CDMFT , which in the toy model 
manifests itself in the site dependence of the bond expectation 
value $\chi_{x}$ is certainly one of them. For example, CDMFT 
predicts a finite temperature phase transition for the one 
dimensional Falikov-Kimball model in the semiclassical limit 
\cite{olivier,thomas}. This is due to the fact that value of the Weiss 
field on the boundary is strongly coupled to the value of the 
propagator on the other boundary and, unfortunately, as discussed 
previously, the error is much larger at the boundary than in the 
bulk. This may not be a serious problem for phases with  broken 
symmetry, but certainly  is in  translationally invariant 
phases.   
One  possible solution of this problem would be  
to modify the self-consistent equations   that express  the 
Weiss field  as a function of the propagator  so as to  use  more 
heavily   bulk values of the propagator,  which have a reduced 
error relatively to the boundary. Clearly, further investigations 
are needed to optimize CDMFT in the light of this point. A related 
issue,  stressed in  ref \cite{biroli,cdmft}, is that the 
lattice self energy in CDMFT is a derived quantity, obtained from 
the cluster self energy entering the CDMFT equations. The lattice 
self-energy is obtained using a matrix w, via the formula (10) in 
ref \cite{biroli}. If w is positive definite, then the lattice 
self-energy is causal, but one can sometimes obtain better 
estimates by using other matrices. For the toy model, the matrix 
w  are not restricted since the self energy is real. The positive 
definitiveness of w is a sufficient, but not a necessary 
condition to maintain causality. A  general constructive way to 
find the best estimator of the lattice self energy (preserving 
causality) for an arbitrary model Hamiltonian is lacking. However, 
an important criterion to follow is trying to extract the 
prediction on the lattice self energy from the bulk values of the 
cluster self-energy in order to minimize the error as discussed 
previously.\\  
None of these issues, however, affect the basic fact that 
expectation values of physical observables which are {\it local} 
(namely defined on a restricted neighborhood in physical space) 
converge as $1 / L_{c}^2$ for DCA as shown in \cite{comment}, while 
for CMDFT they converge exponentially in situations like the one outlined 
for the toy model, when the correlation length is finite, and not 
as $1/L_{c}$ as claimed by K. Aryanpour et al \cite{comment}. By 
exploiting the freedom in the choice of basis, which is inherent 
to the original CMDFT formulation, one can improve convergence of 
observables which become local when the approximation is 
formulated in a different basis set. The problem of convergence 
as function of cluster size at zero temperature or at a quantum 
critical point, or for quantities that are dominated by massless 
excitations,  remains  an open question. However these problems could 
be better  addressed by techniques other than quantum cluster 
methods.\\ 
Finally, most studies can only be done for small clusters, and it 
is important to understand whether the results obtained in small 
clusters are representative of the thermodynamical limit. 
Recent CDMFT studies of the Hubbard model, in one dimension, show 
that while even - odd effects are important, even clusters of 
small size can give very accurate results \cite{civelli} as 
compared with exact Bethe Ansatz results in the thermodynamical 
limit. \\

\begin{acknowledgments} 
We thank O. Parcollet for interesting discussions. G. Biroli acknowledges 
financial support from the French Minister of research under the
ACI project ``Ph{\'e}nom{\`e}nes 
quantiques {\`a} l'{\'e}chelle macroscopique''. 
G. Kotliar is supported by the NSF under grant  DMR -0096462. 
\end{acknowledgments}


\begin{thebibliography}{99} 
\bibitem{biroli} G. Biroli and G. Kotliar , Phys. Rev. B 65, 155112 
(2002) 
\bibitem{cdmft} 
G. Kotliar, S. Y. Savrasov  G. Palsson and G. Biroli Phys. Rev. Lett {\bf 87}, 
186401 (2001). 
\bibitem{dca} 
M. Jarrell, Th. Maier, C. Huscroft, and S. Moukouri, 
Phys. Rev. B {\bf 64}, 1951301 (2001). 
\bibitem{comment}  K. Aryanpour Th. Maier and 
M. Jarrell, {\it The Dynamical Cluster Approximation (DCA) versus the 
Cellular Dynamical Mean Field Theory (CDMFT) 
in strongly correlated electrons systems}, cond-mat/0301460. 
\bibitem{thomas} Th. Maier and M. Jarrell Phys. Rev. B 65, 041104 (2002). 
\bibitem{affleck} 
I. Affleck and B. Marston, Phys. Rev. B {\bf 37}, 3774 (1988). 
\bibitem{olivier}  G. Biroli, O. Parcollet, G. Kotliar, {\it Cluster 
dynamical mean-field theories: Causality and classical limit}, cond-mat/0307587. 
\bibitem{civelli} M. Capone, M. Civelli, V. Kancharla, C. Castellani and G. Kotliar cond-mat 
/0401060 
\end{thebibliography}
\end{document}